\documentclass{PoS}

\usepackage{xspace}
\usepackage[utf8]{inputenc}

\title{Characterization of atmospheric properties at the future sites of the Cherenkov Telescope Array}

\ShortTitle{Characterization of atmospheric properties at CTA sites}

\author{\speaker{Jan Ebr}$^a$, Dušan Mandát$^a$, Miroslav Pech$^a$, Ladislav Chytka$^b$, Jakub Juryšek$^{a,b}$, Michael Prouza$^a$, Petr Janeček$^a$, Petr Trávníček$^a$, Jiří Blažek$^a$, Tomasz Bulik$^c$, Marek Cieslar$^c$, Mariusz Suchenek$^d$, Vincenzo Rizi$^e$, Ermanno Pietropaolo$^e$, Marco Iarlori$^e$, Carla Aramo$^f$, Laura Valore$^g$, Federico Di Pierro$^g$, Piero Vallania$^g$, Davide Depaoli$^g$, Martin Will$^h$, Markus Gaug$^j$ and Lluís Font$^j$ for the CTA consortium and Martin Mašek$^a$, Jiří Eliášek$^a$, Martin Jelínek$^i$ and Sergey Karpov$^a$\\
\llap{$^a$}\textit{FZU -- Institute of Physics of the Czech Academy of Sciences, Czech Republic}\\
\llap{$^b$}\textit{Palacký University, Olomouc, Czech Republic}\\
\llap{$^c$}\textit{Astronomical Observatory, Department of Physics, University of Warsaw, Poland} \\
\llap{$^d$}\textit{Faculty of Physics, University of Warsaw, Poland} \\
\llap{$^e$}\textit{INFN-GSGC-L’Aquila and CETEMPS/DSFC, Università degli Studi dell’Aquila, Italy} \\
\llap{$^f$}\textit{INFN Napoli and Dipartimento di Fisica, Università di Napoli “Federico II”, Italy} \\
\llap{$^g$}\textit{INFN Torino, Italy} \\
\llap{$^h$}\textit{Max Planck Institute for Physics, Munich, Germany} \\
\llap{$^i$}\textit{Astronomical Institute of the Czech Academy of Sciences, Czech Republic} \\
\llap{$^j$}\textit{Departament de Fisica, and CERES-IEEC, Universitat Autonoma de Barcelona, Spain}\\ 

E-mail: \email{ebr@fzu.cz}}

\abstract{Advanced knowledge of the detailed atmospheric properties of both the future sites of the Cherenkov Telescope Array is essential in preparation of the arrival of the first scientific data. Meteorological variables are studied using a dedicated characterization station installed at the southern site in Chile and a wealth of data from existing observatories around the northern site on the La Palma island. Campaigns using radiosondes launched on balloons are foreseen to complement these data in the near future. Cloudiness during the night has been continuously monitored at both sites for several years using All-sky Cameras which assess the presence of clouds based on detection of stars. The integrated aerosol optical depth over the southern site has been measured using a Sun/Moon Photometer since 2016 and the small robotic FRAM telescope since 2017; identical instruments have been deployed at the northern site in autumn 2018. Also in October 2018, the ARCADE Raman lidar (RL) has started to take measurements on routine basis at the northern site, providing data on the vertical profile of the aerosol optical properties (i.e., extinction and scattering) and of the water vapour mixing ratio. We present the data currently available from these instruments from both sites with emphasis on characteristics important for the (future) operation of Imaging Atmospheric Cherenkov Telescopes.}

\FullConference{36th International Cosmic Ray Conference -ICRC2019-\\
July 24th - August 1st, 2019\\
Madison, WI, U.S.A.}

\begin{document}

\begin{figure}[t]
\centering
\includegraphics[width=1.02\textwidth]{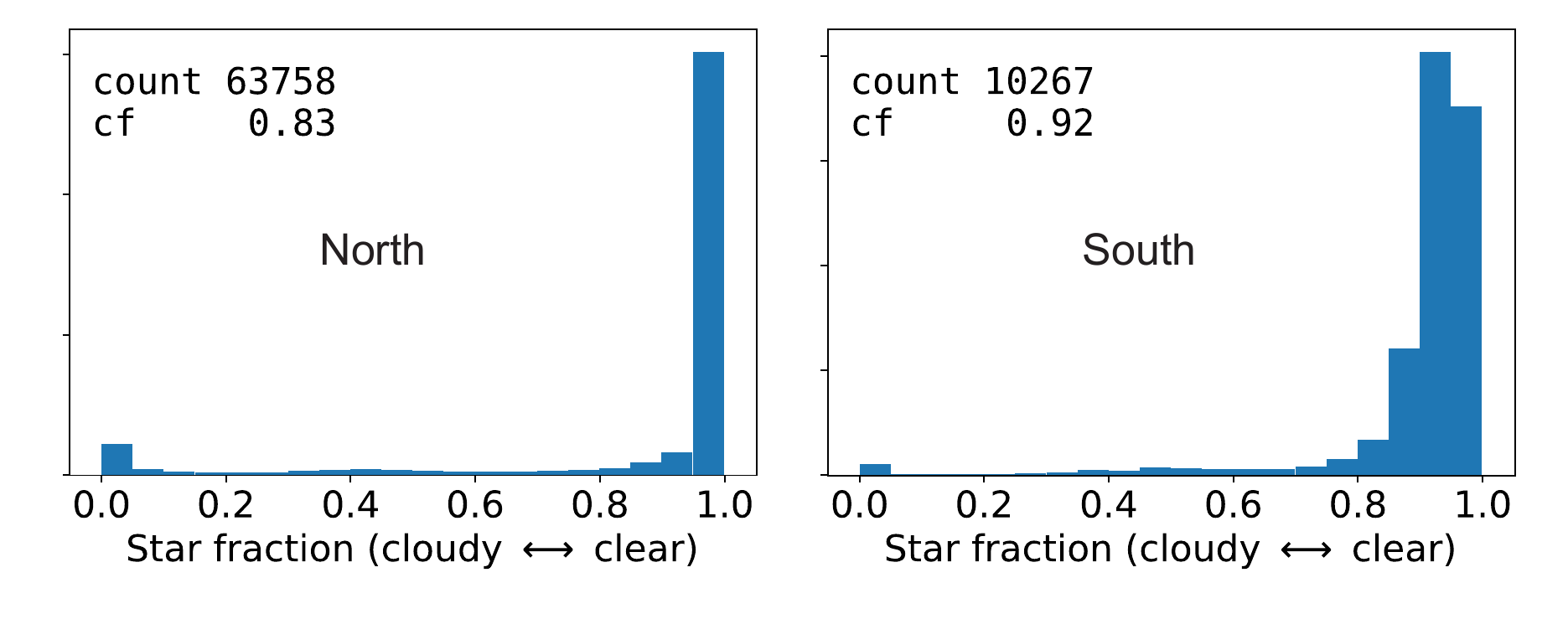}
\caption{\label{fig:ASC} Distribution of the star fraction recorded by the All-sky Cameras at the future CTA sites between 11/2015--12/2018 for the CTA South site and bewteen 10/2015--06/2019 for the CTA North site. The number of individual measurements used (count) is provided as well as the \emph{clear fraction} -- the amount of measurements with less than 20 \% of the sky covered by clouds.}
\end{figure}

\section{Introduction}

The future sites of the Cherenkov Telescope Array (CTA) -- the CTA South site near Cerro Armazones, Chile and the CTA North site on La Palma, Canary Islands, Spain -- are equipped with various devices used for atmospheric characterization. Their brief technical description is mostly provided in \cite{lastICRC} (and references cited therein) and the experience from operating some of them as prototypes of atmospheric monitoring devices to be used during CTA operations is presented in \cite{proto} together with  information on the covered time periods and instrument availability. In this paper, we focus on the results obtained at the sites and provide additional information necessary for proper interpretation of the data.

\section{Cloud coverage}

The cloud coverage above the sites is monitored during nighttime using the All-sky Cameras (ASC) \cite{ASC}, which observe stars over the whole sky simultaneously. Each 30-second exposure is automatically processed and stars detected on the image are compared with the expectation from a catalog. The distribution of the resulting \emph{star fraction} is shown in Fig.~\ref{fig:ASC}; this quantity is, for a full-sky analysis, a very good approximation of the fraction of sky not covered by clouds. For this analysis, only images taken with Moon under the horizon were used, to avoid systematic effects in the distributions -- for operational assessment of cloudiness, all other images can be used as well. For both sites, the \emph{clear fraction} -- the amount of time when less than 20 \% of the sky covered is by clouds and thus the conditions are favorable for observations -- is very large (83 \% and 92 \%).

\begin{figure}[t]
\centering
\includegraphics[width=\textwidth]{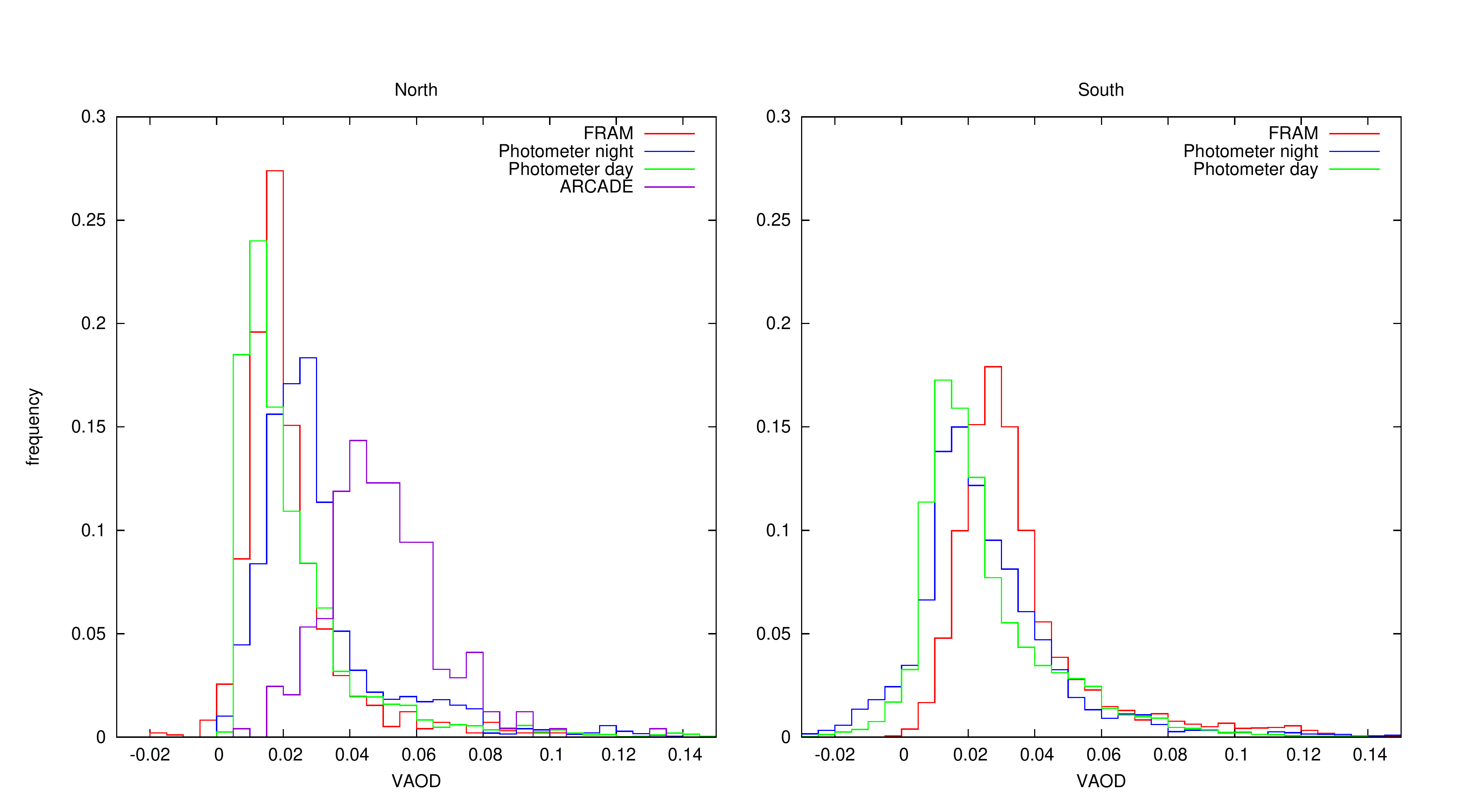}
\caption{\label{fig:vaod} Distribution of the VAOD measured by the different instruments at both sites. In the CTA North site, the ARCADE data span the period 10/2018--04/2019 while FRAM data were taken between 11/2018--05/2019 and Photometer data between 10/2018--06/2019. In the CTA South site, the FRAM data cover 09/2017--12/2018 while the Photometer data come mostly from 08/2017--12/2018 with a small contribution from three months in 2016.}
\end{figure}

\begin{figure}[t]
\centering
\includegraphics[width=0.9\textwidth]{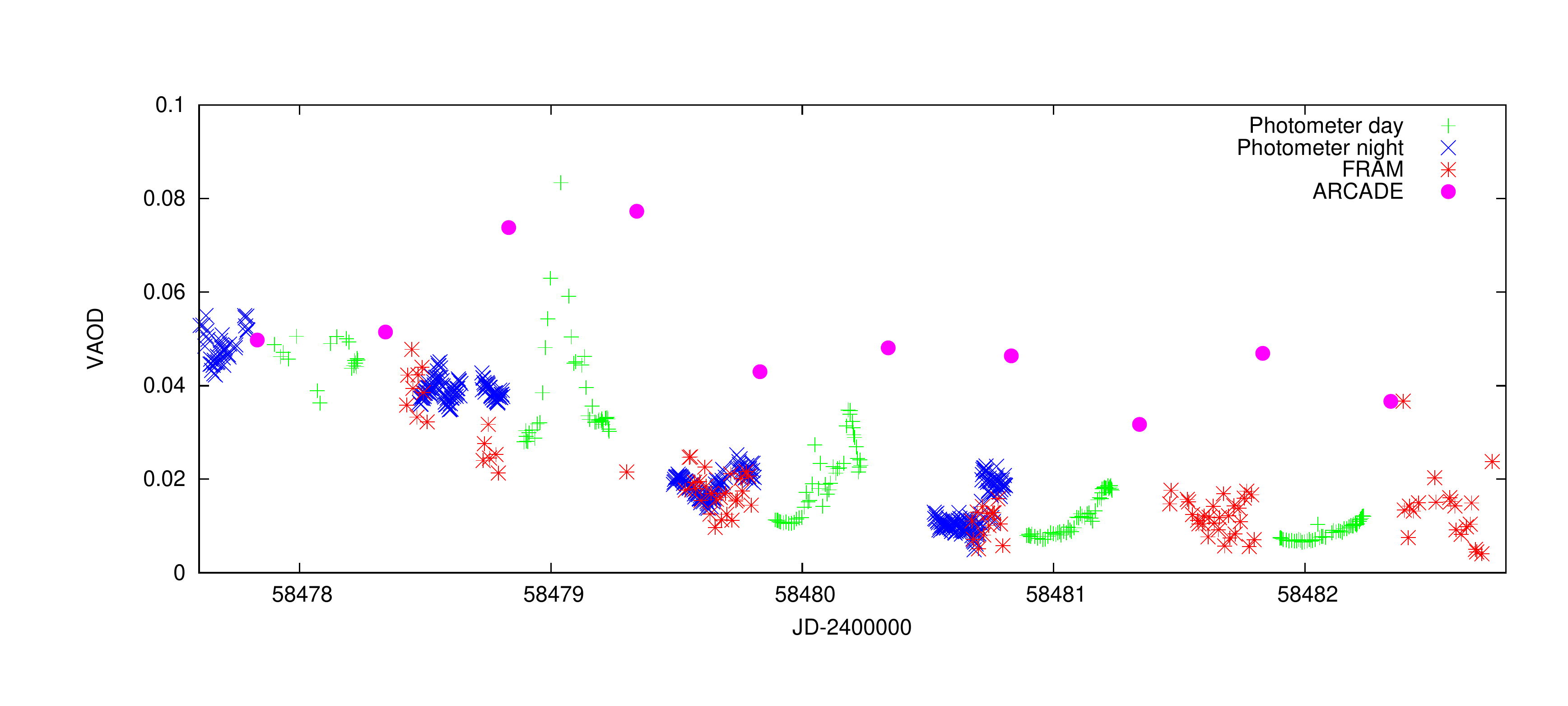}
\caption{\label{fig:vaod-seq} Time sequence of VAOD measurements from the CTA North site in 12/2018.}
\end{figure}

\section{Aerosols}

\subsection{Instruments and uncertainties}

The transparency of the atmosphere above the sites is monitored using the FRAM telescopes \cite{FRAM}, the Sun/Moon Photometers \cite{phot}, and, at CTA North, also with the ARCADE Raman Lidar system \cite{ARCADE} (which also provides other quantities, such as the vertical profiles of back-scatter and extinction coefficients and water vapor profiles). The former two instruments provide only the integral value of the vertical aerosol optical depth (VAOD) through the whole atmosphere, while the Lidar measures the whole vertical profile. For the sake of comparison of the data, we use the VAOD value at 3 km above the ground, which is above the Atmospheric Boundary Layer, where most of the aerosols usually reside, but still close enough to the instrument for a very precise measurement. Clearly, this value is expected to be less or equal to the integral measurement. Due to constraints from astronomical observatories located around the CTA North sites, the measurements are taken only twice a day during twilight. After accumulating data for 15 minutes, the estimated error is less than 0.015 in VAOD at 3 km. The ARCADE system is fixed and thus the measurements are always conducted in the vertical direction and at a single wavelength of 355 nm.

The Sun/Moon Photometer operates in two rather distinct modes. During the day, using the light from the Sun, which is well characterized, stable and abundant, the accuracy of the VAOD measurements is estimated to be better than 0.01 (note that the VAOD is a dimensionless quantity). During the night, if the Moon is at least 40 \% illuminated and at least 10 degrees above the horizon, it can be used as a source as well, but the error is usually estimated to be roughly 0.04. Our analysis shows that this can likely be improved, using stringent quality cuts on the data and removing the residual dependence of the calibration constant of the instrument on lunar phase (which points to improper description of Moon brigtness in the model used) using a fit over many lunar cycles. Considering that the corrected data show, for CTA North, a correlation with the FRAM measurements with a mean shift of 0.0045 and RMS spread of 0.011 (see Fig.~2 in \cite{proto}), it is unlikely for both the devices to possess a much larger uncertainty, barring a correlated systematic shift. However the CTA South data must be interpreted with caution, possibly because of an issue related to rapid fluctuations of the sensitivity of the device. The Photometer measurement is a directed one, in a direction always determined by the position of the Sun or the Moon. It has a large amount of filters available, from which the 440-nm one is most suitable for the comparison with other instruments.

The FRAM telescopes determine the VAOD using photometry of a large number of stars in a large span of altitudes above the horizon and thus seen through varying column depths of air. Assuming horizontal uniformity of aerosols at least in the close vicinity of the instrument (several kilometers), the VAOD can be extracted from a fit of observed light extinction as a function of the altitude above the horizon. This method has been recently significantly improved using laboratory measurements of various properties of the system \cite{FRAM2}; furthermore a thorough study of uncertainties has been done. The statistical uncertainty of a single measurement is 0.002--0.004 and most of the systematic uncertainties (from sources such as the subtraction of molecular extinction, knowledge of the non-linearity of the camera and spectral response of the system etc.) combined are found to be 0.006. A dominant uncertainty of 0.015 is related to the choice of photometric software and requires further examination, but the outlook for the method of choosing the correct model using internal consistency of the data is promising, which would bring the over systematic uncertainty (correlated plus uncorrelated) to around 0.01. The FRAM measurements are performed as a scan between the horizon and the zenith in a chosen azimuth, giving them some directionality, but the extraction of the VAOD assumes a degree of horizontal homogeneity anyway. Using Johnson B filter, the bandpass spans roughly 100 nm, with an effective mean wavelength of 440 nm for typical aerosols.

\subsection {Results}

Fig.~\ref{fig:vaod} shows the comparison of the VAOD distribution measured by the different instruments. The problematic behavior of the CTA South Photometer is obvious from the presence of spurious negative VAOD values. The difference between the ARCADE data and other measurements is significant compared to the stated uncertainties and shall be investigated further, given that the Raman Lidar is traditionally considered to be the reference instrument for aerosol measurements -- however, it is important to note, that data from all instruments should be considered preliminary, pending improvements in the analyses.  

Another point to note is that the periods covered slightly differ and even if the devices were covering the same periods, they did not take data at the same time and in the same direction. In particular, the FRAM measurement requires a full undisturbed swath of clear sky for the VAOD fit, which means that in periods of scattered clouds, it does not produce VAOD data, while for the Photometer, only the direction of the Moon needs to be cloud-free and for the ARCADE measurement, any clouds higher than 3 km above observation level are irrelevant. This dependence on weather may reflect in a bias in the distribution, as aerosol concentration can easily be correlated with weather. To alleviate this issue, an algorithm that selects the direction for the FRAM observations based on real-time ASC data in order to avoid clouds as much as possible is being implemented. Moreover, the FRAM has to specifically avoid the direction of the Moon on the sky (as direct moonlight saturates the camera), where the Photometer is taking data, so the effect any inhomogeneity is exacerbated.

\begin{figure}[t]
\centering
\includegraphics[width=\textwidth]{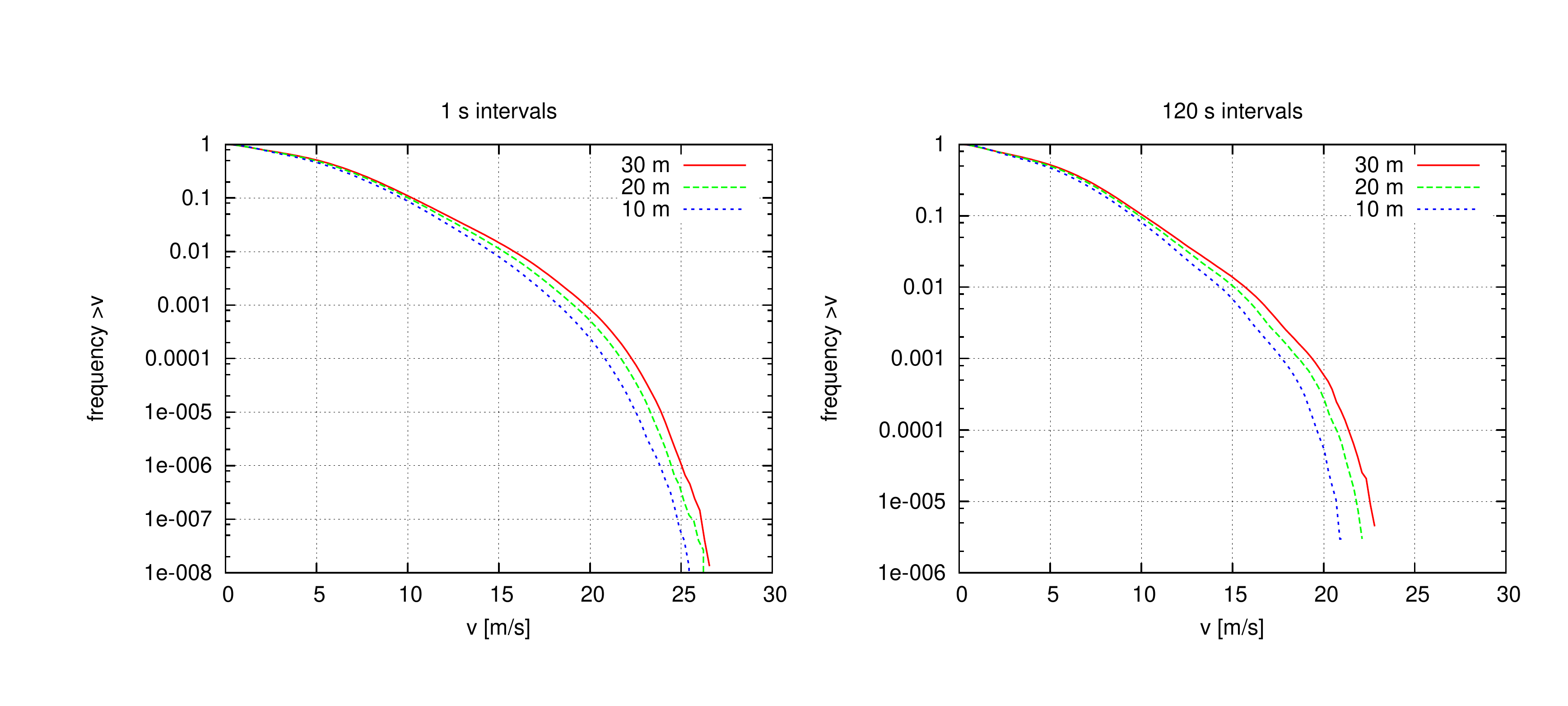}
\caption{\label{fig:wind} The fraction of wind speed above a certain velocity measured at three different heights at the CTA South site between 02/2016--07/2018. The left panels shows data with 1-second granularity, while in the right panel, averages over 120 seconds are used to smooth out wind gusts. Over 75 million 1-second measurements have been recorded in this time period. }
\end{figure}

\begin{figure}[t]
\centering
\includegraphics[width=\textwidth]{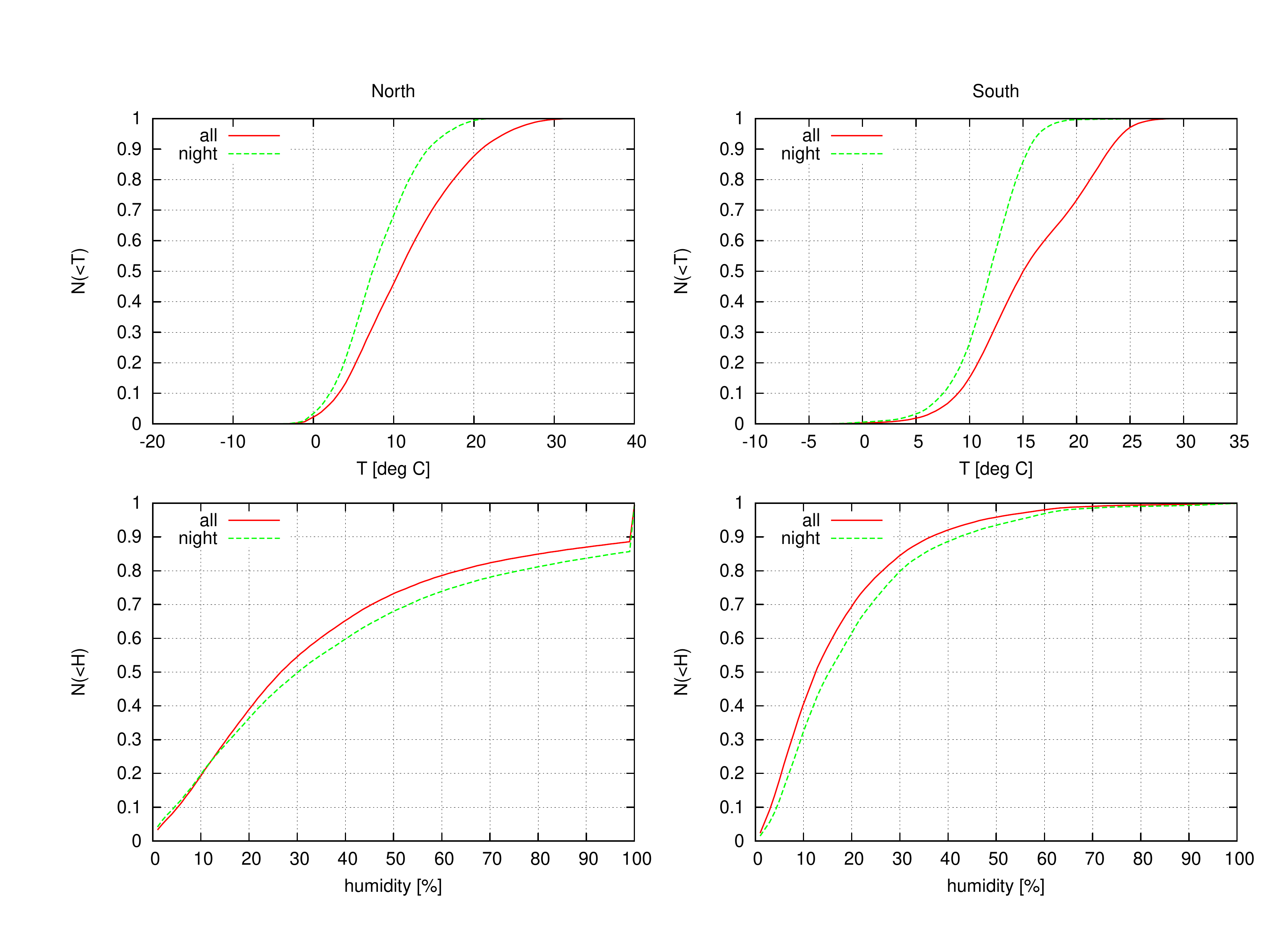}
\caption[dummy]{\label{fig:meteo} Cumulative distributions of temperature (top panels) and relative humidity (bottom panels) observed at the future CTA sites. For CTA South, data from the weather station installed on a 10-metier mast as a part of the site characterization station are used, for CTA North, the data come from a weather station operated by the MAGIC Collaboration on the roof of the MAGIC Counting house \protect\footnotemark. Additionally to distributions of all measured values, the values for nighttime periods (defined by nautical twilight) are shown separately. The data from the CTA South site cover the period between 05/2015--12/2018 and the data from CTA North the period between 11/2012--02/2019. }
\end{figure}

Fig.~\ref{fig:vaod-seq} shows a time sequence of VAOD observations in a selected time period in 12/2018. The typical trend of increase during the day and then decrease during the night is obvious as is a slow change of conditions over several days. The daytime Photometer measurements are the most reliable and thus seeing the nighttime measurements from Photometer and FRAM interpolate well between days is encouraging. However the data also show that the nighttime and daytime behavior of aerosols can be significantly different and thus the improvements in data analysis for the nighttime data are worth pursuing for the purpose of site characterization.

From the point of view of the future CTA Observatory, both Sun/Moon Photometer and FRAM data suggest that both sites are exceptionally clean, with very consistent aerosol contents over time, while the ARCADE measurements could be interpreted as slightly less favorable for the northern site. 

\footnotetext{An article describing this weather station and its results is currently in preparation -- M. Gaug, O. Blanch, L. Font, M. Doro and J. Zapatero, \emph{15 years of MAGIC weather station data}.}

\section{Meteorological observations}

A large amount of meteorological observations on both sites are available thanks to the dedicated site characterization station deployed at the CTA South site and to several major astrophysical observatories close to the CTA North site, among all the MAGIC telescopes located directly within the future CTA area. From the monitored variables,  possibly the most impactful on the construction and operation of the CTA Observatory is the measurement of wind in three heights above ground at the CTA South site as shown in Fig.~\ref{fig:wind}. No wind gust over 27 m/s has been recorded at the site and the differences between the different heights are small. Further values of interest for the safe and efficient operations of telescopes are the distributions of temperature and humidity (Fig.~\ref{fig:meteo}) -- note that the CTA South site is located in one of the driest deserts of the planet and thus the humidity is consistently low, while at the CTA North site there are periods of high humidity when the upper reaches of Roque de los Muchachos are engulfed by clouds. The temperature differences between day and night are also significantly more pronounced in the desert of the CTA South site. 

\section*{Acknowledgements}

We gratefully acknowledge financial support from the agencies and
organizations listed here: \href{http://www.cta-observatory.org/consortium\_acknowledgments}{www.cta-observatory.org/consortium\_acknowledgments}, in particular by Ministry of Education, Youth and Sports of the Czech Republic (MEYS) under the projects MEYS LM2015046, LTT17006 and 
EU/MEYS CZ.02.1.01/0.0/0.0/16\_013/0001403 and European Structural and Investment Fund and MEYS (Project CoGraDS -- CZ.02.1.01/0.0/0.0/15\_003/0000437), CETEMPS-UNIVAQ-Italy and NCN, Poland. The installation of the devices on La Palma would not be possible without the support from the MAGIC Collaboration. This work was conducted in the context of the CTA Consortium.

\end{document}